\documentclass[%
 aps,
 sd,%
 amsmath,amssymb,
 reprint,%
]{revtex4-1}
\usepackage{graphicx}
\usepackage{dcolumn}
\usepackage{bm}
\usepackage{tabularx}
\usepackage{epstopdf}
\graphicspath{{./}}
\usepackage{subfigure}
\usepackage{gensymb}
\usepackage{xspace}	
\newcommand{\degg}{$^{\circ}$\xspace}
\usepackage{multirow}
\usepackage{footmisc}
\usepackage{chngcntr}

\newcommand{\ta}{TaSe$_3$\xspace}

\newcommand{\zr}{ZrTe$_3$\xspace}
\newcommand{\wmk}{{Wm$^{-1}$K$^{-1}$}\xspace}
\newcommand{\cmu}{{cm$^{-1}$}\xspace}
\newcommand{\xh}{$\hat{x}$\xspace}

\newcommand{\kx}{\kappa_{xx}}
\newcommand{\ky}{\kappa_{yy}}
\newcommand{\kz}{\kappa_{zz}}
\newcommand{\kcc}{\kappa_{cc}}
\newcommand{\kcp}{\kappa_{cp}}

\newcommand{\av}{{\bf a}}
\newcommand{\bv}{{\bf b}}
\newcommand{\cv}{{\bf c}}

\usepackage[unicode=true,
 bookmarks=false,
 breaklinks=false,pdfborder={0 0 1},colorlinks=true] 
 {hyperref}
\hypersetup{
 pdftex,pdfstartview=FitV,linkcolor=blue,citecolor=blue,urlcolor=blue}

\begin{document}

\preprint{AIP/123-QED}

\title[]{Thermal conductivity of the quasi-1D materials \ta and \zr}

{
\makeatletter
\def\frontmatter@thefootnote{%
 \altaffilletter@sw{\@fnsymbol}{\@fnsymbol}{\csname c@\@mpfn\endcsname}%
}%
\makeatother

\author{Topojit Debnath}
\email[]{tdebn001@ucr.edu}
\affiliation{Department of Electrical and Computer Engineering, University of California, Riverside, CA 92521, USA}

\author{Bishwajit Debnath}
\email[]{Current affiliation : Intel Corporation, Hillsboro, Oregon}
\affiliation{Department of Electrical and Computer Engineering, University of California, Riverside, CA 92521, USA}

\author{Roger K. Lake}
\email[]{Corresponding author: rlake@ece.ucr.edu}
\affiliation{Department of Electrical and Computer Engineering, University of California, Riverside, CA 92521, USA}


\begin{abstract}
The high breakdown current densities and resilience to scaling
of the metallic transition metal trichalcogenides \ta and \zr 
make them of interest for possible interconnect applications,
and it motivates this paper of their thermal conductivities and phonon properties. 
These crystals consist of planes of strongly bonded one-dimensional chains more weakly bonded to neighboring chains.
Phonon dispersions and the thermal conductivity tensors are calculated using
density functional theory combined with an
iterative solution of the phonon Boltzmann transport equation. 
The phonon velocities and the thermal conductivities
of \ta are considerably more anisotropic than those of \zr.
The maximum longitudinal-acoustic velocity in \zr occurs in the cross-chain direction, and this is consistent with the
strong cross-chain bonding that gives rise to large Fermi velocities in that direction.
The thermal conductivities are similar to those of other metallic two-dimensional
transition metal dichalcogenides.
At room temperature, a significant portion of the heat is carried by the optical modes.
In the low frequency range, the phonon lifetimes and mean free paths in \ta are considerably shorter
than those in \zr.
The shorter lifetimes in \ta are consistent with the presence of lower frequency optical branches and 
zone-folding features in the acoustic branches that arise due to the doubling of the \ta unit cell within the plane.
\end{abstract}

\maketitle

\section{Introduction} 
The transition metal trichalcogenides (TMTs)
have a quasi-one-dimensional (quasi-1D) crystalline structure that can give rise to
quasi-1D behavior of the electronic and phononic properties 
\cite{Bullett_CDWs_in_TMTs_JPC79,TX3_Review_JMatSci92,2017_Joshua_review}.
The lattice structures of TMTs
consist of a transition metal (\textit{M}) atom, contained at the 
center of a prism of chalcogen (\textit{X}) atoms \cite{Crystal_Structs_MX3s_75}.
The \textit{MX}$_3$ prisms create strongly \textit{M-X} covalently bonded chains arranged 
side-by-side via longer, weaker \textit{M-X} bonds, and planarly stacked with even weaker van der Waals type \textit{X-X} bonds which effectively makes these materials two-dimensional 
layers of quasi-1D chains \cite{1981_endo_photoelectron},
giving them the apt name of quasi-1D materials.
%
%
The crystalline anisotropy gives rise to 
directional anisotropy in electrical, optical, phononic, and thermal 
properties\cite{1984_Sellmyer_electrical, 2015_Yang_1D, dai2015titanium,
TiS3_Gomez_SciRep16, pant2016strong, dai2016group, kong2017angle}. 

The low dimensionality of the \textit{MX}$_3$ materials makes them particularly susceptible to 
multiple phase transitions such as superconductivity 
\cite{1977_Samboongi_superconductivity,
1978_Haen_low_T_phase,
1978_yamamoto_superconducting,
ZrTe3_SC_CDW_PRB12} 
and charge density wave (CDW) formation
\cite{Gruner_RMP88,
Gruner_CDWs,
Monceau_Review_AdvPhys12,
Hi_f_quantum_effects_CDWs13}. 
The latter has motivated extensive, prolonged research into CDW transitions 
\cite{
Fleming_NbSe3_CDW_PRB78,
Bullett_CDWs_in_TMTs_JPC79,
MItkis_TaS3_JPCM90,
2017_Zybtsev}, 
sliding 
\cite{Zettl_Gruner_Onset_CDW_cond_PRB82,Zettl_Switching_CDW_I_PRB88}, 
dynamics \cite{Bardeen_PRL82,
Zettl_Gruner_CDW_Trans_f_TaS3_PRB82,
Thorne_Tucker_Lyding_Bardeen_ac_dc_PRB87,
Zant_NDR_PRL01,
GHz_CDW_APL09},
 dimensional scaling 
\cite{
Zotov_finite_size_review04,
McCarten_finite_size_PRB92,
Meso_NbSe3_wires,
Zant_01,
Thorne_Xover_2D_1D_NbSe3_PRB04,
Inagaki_APL05,
NWs_NRibbons_CDW_NbSe3_NL05,
Discrete_NbSe3_CDW_PRB11} 
in the metallic TMTs,
and their use for device 
applications \cite{Bhattacharya_patent86,CDW_FET_PRL95,CDW_Capacitor_Patent,fs_data_storage_CDW_APL02}. 
The Fermi surface nesting, CDW, and their signatures in the phonon spectrum of \zr are a topic of 
ongoing interest 
\cite{1998_FSN_ZrTe3_Stowe_JSSChem,
ZrTe3_CDW_transition_63K_Felser98,
ZrTe3_e-properties_Felser98,
2005_CDW_ARPES_ZrTe3_PRB,
2009_Kohn_Anomaly_PRL,
2009_FS_Splitting_ZrTe3_HBerger_PRB,
2015_ep_coupling_ZrTe3_PRB,
2015_CDW_2_SC_ZrTe3_PRB,
2018_ZrTe3_Ni_interc_PRB,
2019_Disorder_Quench_CDW_ZrTe3_PRL,
2020_Pristine_Disorder_CDW_ZrTe3_NCom}. 
Inelastic x-ray scattering revealed a Kohn anomaly in the transverse acoustic (TA) phonon of \zr at the 
CDW wavevector 
for temperatures up to 292 K \cite{2009_Kohn_Anomaly_PRL}. 
%
%
As the temperature was reduced to and below $T_{\rm CDW}$,
partial Fermi surface splitting was observed by angle-resolved photoelectron spectroscopy \cite{2005_CDW_ARPES_ZrTe3_PRB,2009_FS_Splitting_ZrTe3_HBerger_PRB}.

There has been a recent resurgence of interest in 1D and quasi-1D materials, in part 
motivated by, and evolving from, the intense interest in two-dimensional (2D) van der Waals materials.
Reference \cite{2020_TMTs_Review_RSC_Adv} provides a review of the more recent work on the 
transition metal trichalcogenides.
The direct gap semiconductor TiS$_3$ has received much attention after it was exfoliated to few layer 
thicknesses and shown to have a high photo-response with a bandgap of 1.1 eV \cite{TiS3_NRs_Photoresponse_AdvOptMat14}.
Investigations of few layer and nanoribbon TiS$_3$ transistors followed, and experimentally extracted mobilities
were in the range of 20-70 cm$^2$V$^{-1}$s$^{-1}$ 
\cite{TiS3_FET_C-Gomez_AdvMat15,
TiS3_FET_Sinitskii_NScale15,
TiS3_Gomez_SciRep16,
TiS3_Dowbden_Contacts_APL19,
TiS3_FET_MIT_Bird_ACSNano19}. 
BN encapsulation of 26 nm thick TiS$_3$ resulted in two and four terminal room-temperature 
mobilities of 54 and 122 cm$^2$V$^{-1}$s$^{-1}$, respectively \cite{BN_TiS3_2DMat19}.
The properties of the \textit{MX$_3$} materials with M = Ti, Zr, Hf, and X = S, Se, Te,
have been investigated with density functional theory (DFT) to determine electronic structure, bandgaps, specific heats, and elastic 
constants \cite{MX3s_Joubert_EurPhysJB15}.
Exploiting the anisotropy inherent in TMTs 
has been proposed for application in next generation electronics, contacts, polarizers and photo-detectors 
\cite{zhang2016tunable, yuan2015polarization}.

The metallic TMTs have also received renewed attention after it was discovered that
the breakdown current densities of \ta ($\sim 10$ MA/cm$^2$ where MA represents mega-amperes) and \zr ($\sim 100$ MA/cm$^2$) are higher than that of Cu
\cite{2016_Balandin_TaSe3,2018_Balandin_ZrTe3}. 
Furthermore, in contrast to Cu wires, the resistivity of \ta did not degrade as the 
cross-sectional dimensions were scaled down to 10 nm \cite{TaSe3_10nm_Bartels_NLett19}.
This was attributed to the single crystalline nature of the nanowires and the self-passivation
of the surfaces that eliminate grain-boundary and surface roughness scattering.
Because of these properties, metallic TMTs were proposed for local interconnect applications in nano-scaled
electronics \cite{2016_Balandin_TaSe3,TaSe3_10nm_Bartels_NLett19}.

\begin{figure*}[!t]
\centering \includegraphics[height=4in]{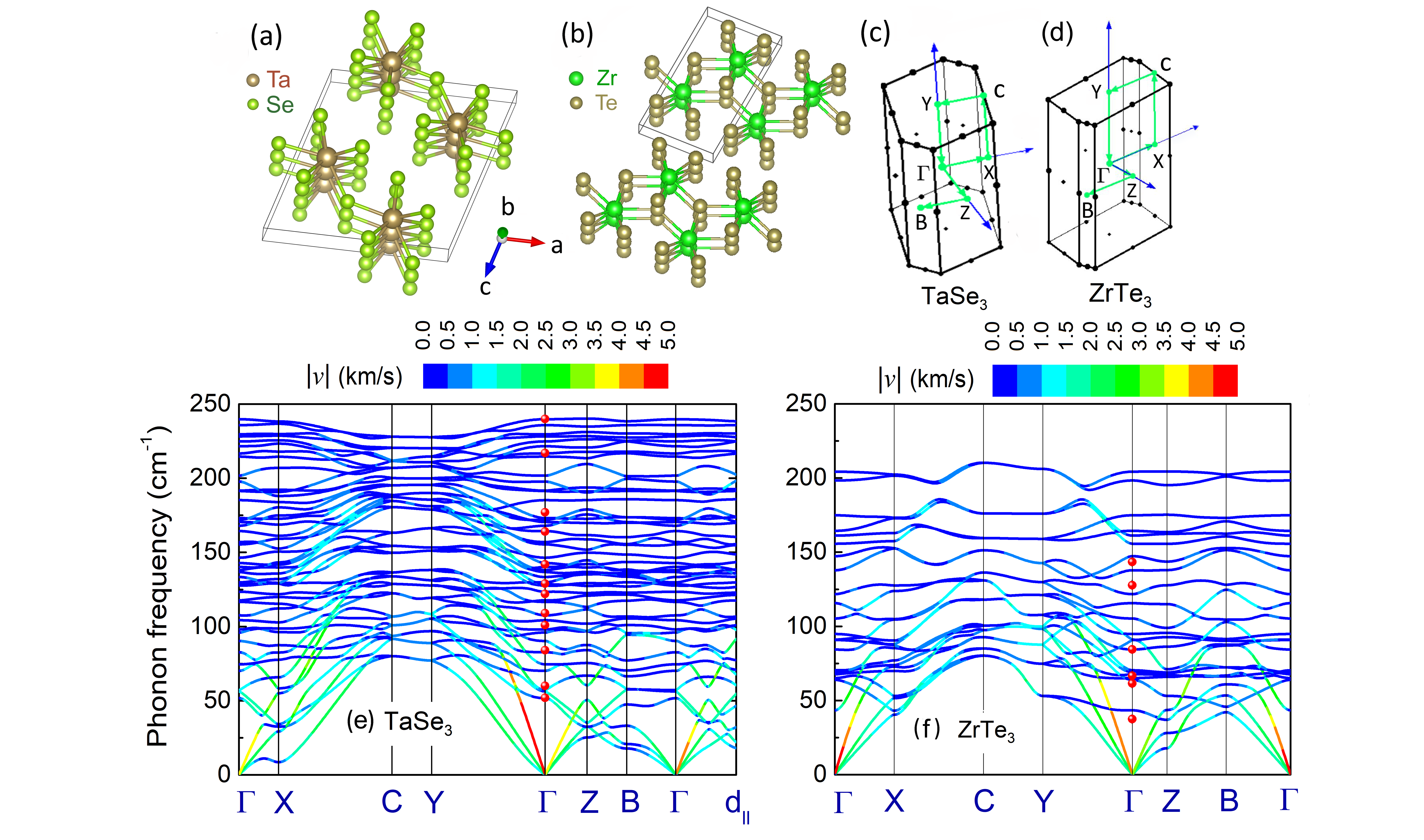}
\caption{
Crystal structures and unit cells of (a) \ta, (b) \zr, and
the corresponding Brillouin zones (BZs) for (c) \ta and (d) \zr.
The wires grow along the $b$ axis. 
Lattice vectors $a$ and $b$ lie along the $x$ and $y$ axes, respectively.
Lattice vector $c$ lies at angles $\beta$ with respect to the $x$-axis with
values of $106.36^\circ$ and $97.95^\circ$ for
\ta and \zr, respectively. 
Phonon dispersions of bulk (e) \ta and (f) \zr.
The color indicates the magnitude of the group velocity as given in the color bar.
The directions in the Brillouin zone are shown in (c) and (d).
The red spheres indicate experimental Raman peaks at 300 K \cite{1982_Levy_raman,Zwick_1980}.
For both crystals, $\Gamma-Y$ is the chain direction. 
For \ta, $\Gamma-B$ is the cross-plane direction, 
and $\Gamma-d_{||}$ is the cross-chain direction, 
where ${\bf d}_{||} = ( {\bf a}^* + 0.8536 {\bf c}^* ) / 2$. 
For \zr, $\Gamma-Z$ is the cross-plane direction. 
$\Gamma-X$ is $8.0^\circ$ off from the cross-chain direction, however the differences in dispersions
between the true cross-chain direction and $\Gamma - X$ are negligible.
}
\label{fig:TaSe3_NbS3_unitcell}
\end{figure*}

Since the metallic TMTs can carry record current densities and have been proposed for interconnect
applications,
it is necessary to also understand their thermal transport properties.
%
A secondary motivation to explore the thermal transport properties of low-dimensional materials is
their potential of exhibiting higher thermoelectric efficiency \cite{1999_Dresselhaus_lowD_thermoelectric}.
Recently, a high thermoelectric figure of merit was predicted for monolayer ZrSe$_3$ resulting from 
a high power factor and relatively low thermal conductivity \cite{ZrSe3_ZT_ACSApplMatInt18}.
%
%
%
%
%

Although the electronic structure and the electronic transport properties of 
quasi-1D materials have been extensively studied, 
their phononic and thermal transport properties have received less attention.
Recently, a thermal conductivity of 7 Wm$^{-1}$K$^{-1}$ was measured in bulk
polycrystalline ZrTe$_3$ \cite{Hooda_2019}.
A DFT investigation of monolayer ZrSe$_3$ found room temperature thermal conductivities parallel to the chain direction of 
8 Wm$^{-1}$K$^{-1}$ and perpendicular to the chain direction of 
3 Wm$^{-1}$K$^{-1}$ [\onlinecite{ZrSe3_ZT_ACSApplMatInt18}].
A recent study of TiS$_3$ found room temperature thermal conductivities of 
5.78 Wm$^{-1}$K$^{-1}$ and 2.84 Wm$^{-1}$K$^{-1}$ along the chain and interchain directions, 
respectively \cite{TiS3_Heat_OpPhonons_NLett20}.

In this paper, we calculate the phonon modes of \ta and \zr,
using DFT. 
We determine the anisotropic velocities of the acoustic branches along the high-symmetry directions,
and we compare the frequencies of the optical modes to experimental Raman frequencies and determine the
associated mode displacements.
We then solve the phonon Boltzmann transport equation (PBTE)
to determine the lattice thermal conductivities
and to investigate the phonon lifetimes and mean free paths in these two quasi-1D materials.

\section{Crystallographic Structures and Properties}
\label{sec:crystal_structures}
Because the TMTs consist of planes of 1D chains of strongly bonded trigonal MX$_3$ prisms
with each chain bonded to its neighbor chain through longer, weaker
M-X bonds and each plane bonded to its neighbor plane through even weaker
van der Waals type bonds, we define three different directions:
chain, cross-chain (cc), and cross-plane (cp). 
The chain direction is parallel to the 1D chains and along the ${\bf b}$ lattice vector
for both crystals as shown in Fig. \ref{fig:TaSe3_NbS3_unitcell}.
The cross-chain direction is perpendicular to the 1D chains and within the same van der Waals plane,
and the cross-plane direction is perpendicular to the van der Waals planes.

\ta and \zr are metallic with monoclinic crystal structures belonging to space group P2$_1$/m. 
Both crystals have inversion symmetry 
so that the phonon modes at $\Gamma$ have either even or odd parity.
The experimentally determined lattice constants of \ta are
$a = 10.411$ \AA, $b = 3.494$ \AA, $c = 9.836$ \AA, and $\beta = 106.36$\degg; 
and those of \zr are
$a = 5.895$ \AA , $b = 3.926$ \AA , $c = 10.104$ \AA, and $\beta = 97.93$\degg as shown in 
Fig. \ref{fig:TaSe3_NbS3_unitcell}(a,b).\cite{1965_Crystal_structure} 
The van der Waals gaps are visible, and the interchain metal-chalcogen bond is longer than the
intrachain metal-chalcogen bond.
The unit cell of \ta contains 16 atoms, and the unit cell of \zr contains 8 atoms.
%
%
The electrical conductivity along the chain direction of \ta at room temperature 
is reported to be $\sim$1.7 $\times$ 10$^{5}$ ($\Omega$m)$^{-1}$
[\onlinecite{1978_Haen_low_T_phase}] and it remains metallic down to liquid He temperatures
\cite{1977_Samboongi_superconductivity, 1978_Haen_low_T_phase}.
\zr undergoes a CDW phase transition at $T_{CDW}$ = 63 K \cite{ZrTe3_CDW_transition_63K_Felser98,2002_NAKAJIMA}.
For that reason, we limit our thermal conductivity calculations to temperatures 100 K and above.

The structural and electronic properties of ZrTe$_3$ have been investigated extensively both experimentally and
theoretically \cite{takahashi_1984,ZrTe3_CDW_transition_63K_Felser98,ZrTe3_e-properties_Felser98}.
%
The electrical conductivity of ZrTe$_3$ is highest in the {\em cross-chain} direction 
(along the direction of the $\av$ lattice vector),
and, depending on the experimental work, it is a factor of 1.4 to 1.9 larger 
than the conductivity in the {\em chain} direction
(along the $\bv$ lattice vector) \cite{takahashi_1984,ZrTe3_e-properties_Felser98}.
At $T=300$ K, experimental values of the electrical conductivity measured along the $a$-axis
are $5.6 \times 10^5$ ($\Omega$m)$^{-1}$ [\onlinecite{takahashi_1984}]
and $1.3 \times 10^6$ ($\Omega$m)$^{-1}$ [\onlinecite{ZrTe3_e-properties_Felser98}]; 
and the values measured along the $b$-axis are
$4.0 \times 10^{5}$ ($\Omega$m)$^{-1}$ [\onlinecite{takahashi_1984}] and 
$6.9 \times 10^{5}$ ($\Omega$m)$^{-1}$ [\onlinecite{ZrTe3_e-properties_Felser98}]. 
This is a result of Te-Te $\sigma$ and $\sigma^*$ bands formed from Te $p_x$ orbitals. 
The intrachain Te-Te distance is 2.80 {\AA}, and the cross-chain distance is 3.10 {\AA} allowing for
Te-Te bonding that results in highly dispersive electronic bands.
We will see a similar anisotropy in the acoustic phonon velocities with the highest velocity also
occuring in the {\em cross-chain} direction.

To investigate the effect of crystal anisotropy, we calculate
the phonon dispersions and the acoustic phonon velocities along the chain, cross-chain, and
cross-plane directions.
The chain direction always corresponds to the ${\bf b}^*$ reciprocal lattice
vector and the $\Gamma - Y$ paths in the Brillouin zones shown in Fig. \ref{fig:TaSe3_NbS3_unitcell}(c,d).
For \zr, the cross-chain direction also corresponds to the ${\bf a}$ lattice vector in 
Fig. \ref{fig:TaSe3_NbS3_unitcell}(b) and the \xh direction in Cartesian coordinates.
The cross-plane direction is perpendicular to ${\bf a}$ and therefore parallel to 
the ${\bf c}^*$ reciprocal lattice vector.
Thus, for \zr, 
the phonon dispersion in the cross-plane direction is along the $\Gamma - Z$
path in the Brillouin zone shown in Fig. \ref{fig:TaSe3_NbS3_unitcell}(d).
The closest high symmetry line in the Brillouin zone to the cross-chain direction (\xh) is
$\Gamma - X$ and this differs from \xh by $8.0^\circ$ for \zr.
In terms of the reciprocal lattice vectors, the \xh direction for 
ZrTe$_3$ is ${\bf a}^* - 0.2353 {\bf c}^*$.
%
%

For \ta, the ${\bf a}$ and ${\bf c}$ lattice vectors have both cross-chain and cross-plane components,
and the ${\bf a}$ lattice vector points in the \xh direction.
The cross-chain direction is in the direction
of the sum of the lattice vectors ${\bf a} + {\bf c}$.\cite{Anisotropic_Compressibilities_TaSe3_82} 
The cross-plane direction is perpendicular to ${\bf a} + {\bf c}$ and therefore in the direction of
${\bf b} \times ({\bf a} + {\bf c}) \propto ({\bf a}^* - {\bf c}^*)$ 
where ${\bf a}^*$ and ${\bf c}^*$ are the reciprocal lattice vectors in the $x-y$ plane.
Thus, for \ta, the phonon dispersion in the cross-plane direction is along the $\Gamma - B$
path in the Brillouin zone shown in Fig. \ref{fig:TaSe3_NbS3_unitcell}(c).
The closest high symmetry line in the Brillouin zone to the cross-chain direction 
is along ${\bf a}^* + {\bf c}^*$ which runs between $\Gamma$ and the 
corner of the $X$ and $Z$ faces in Fig. \ref{fig:TaSe3_NbS3_unitcell}(c).
This differs from the true cross chain direction by $3.4^\circ$.
The true cross chain direction is ${\bf a}^* + 0.8536 {\bf c}^*$.

\section{Methods}
Structural optimization of each material
is performed using density functional theory 
with projector-augmented-wave (PAW) method \cite{PAW} 
and Perdew-Burke-Ernzerhof (PBE) exchange
correlation functionals \cite{PBE}, as implemented in the Vienna ab initio simulation
package (VASP) \cite{1993_VASP, 1996_VASP}.
The van der Waals interactions are included
by semi-empirical correction of Grimme-D2 \cite{grimmevdW}.
Converged Monkhorst-Pack grids of 9$\times$9$\times$3 are used for \ta and \zr.
All structures are relaxed until the forces on each atom is 
less than 10$^{-5}$ eV/\AA\, and the energy convergence reaches 10$^{-8}$ eV. 
The relaxed lattice constants of \ta and \zr are within 
1\% of the experimentally reported values.
The magnitudes and angles are provided in the Appendix along with the calculated electronic dispersion.

\begin{table*}[!t]
\centering
\def\arraystretch{1.2}
\caption{Velocity (m/s) 
of LA and TA modes near $\Gamma$ along four high symmetry directions and the exact cross-chain direction
labeled as ${\bf d}_{cc}$. 
The chain direction is $\Gamma-Y$. 
The cross-plane direction is $\Gamma-B$ for \ta and $\Gamma-Z$ for \zr.
The first row for each mode corresponds to the velocities at $\Gamma$.
If the maximum velocity occurs at finite frequency, its value and corresponding frequency (cm$^{-1}$)
are provided in the second row.
\vspace{0.10cm}
A `-' in the second row indicates that the maximum velocity occurs at $\Gamma$.
}
\label{tab:velocities}
\setlength{\tabcolsep}{21pt}
\begin{tabular}{c  c  c  c  c  c  c}
\hline
\hline
Material & Mode & $\Gamma - X$ & $\Gamma-Y$ & $\Gamma - Z$ & $\Gamma - B$ & ${\bf d}_{cc}$\\
\hline
\mbox{ } & LA & 3962  & 4956  & 3814  & 2342  & 4105  \\
\ta & TA$_1$ & 1558 & 1256 & 2000 & 1467  & 1824  \\
\mbox{ } & \mbox{ } & -  & 2143 (35.6) & - & - & -\\
\mbox{ } & TA$_2$ & 2374 & 2294  & 2446   & 1963  & 2481  \\ 
\vspace{0.15cm}
\mbox{ } & & - & 2409 (38.0) & - & - & - \\ 
\mbox{ } & LA	& 4723	& 4343  &	3256  & 4686  & 4734 \\
\zr & TA$_1$	& 1618 	& 1629  &	1562  & 1634  & 1511 \\
\mbox{ } & 	& -	& 1829 (23.1) &	- & 1651 (13.5) & - \\
\mbox{ } & TA$_2$	& 2127 	& 2406  &	2257  & 2391  & 2081 \\
    & 	    & -	& 2563 (37.4) &	- & - & -\\
\hline
\hline
\end{tabular}
\end{table*}

To obtain the phonon frequency dispersion and other thermodynamics properties, 
the second-order (harmonic) interatomic force constants (IFCs) are required.
The second-order IFCs are calculated using the finite-displacement, supercell approach as implemented in 
PHONOPY \cite{phonopy,phonopy_2015}. 
For the phonon dispersion of \ta and \zr, 
a supercell size of 2$\times$2$\times$2, 
has been used, with a K-point grid of 
2$\times$6$\times$2 and 6$\times$6$\times$4, respectively. 

The thermal conductivity tensor is 
calculated from the phonon Boltzmann transport equation,
within the three-phonon scattering approximation as implemented within ShengBTE,
\cite{ShengBTE, 2015_Mingo_PRB_YbFeSb}
\begin{equation}
\kappa_{\alpha\beta} = \frac{1}{k_B T^2 NV}\sum_\lambda f_\lambda^0(f_\lambda^0+1) 
						(\hbar \omega_\lambda)^2 v_{\lambda}^{\alpha} F_{\lambda}^{\beta}. 
\label{eq:kappa}
\end{equation}
In Eq. (\ref{eq:kappa}), $\omega_\lambda$ is the phonon energy of each phonon mode $\lambda$, 
$f_{\lambda}^0$ is the equilibrium 
Bose-Einstein distribution of that mode and 
$v_\lambda$ is the group velocity. 
$\lambda$ represents both the phonon branch index $p$ and wave vector $\mathbf{q}$.
$V$ is the volume and $N$ is the number of $\mathbf{q}$ points in the irreducible Brillouin zone.
The quantity $F_\lambda^\beta = \tau_\lambda^0 ( v_\lambda^\beta + \Delta_\lambda^\beta )$
where $\tau_\lambda^0$ is the lifetime in the relaxation-time approximation (RTA), and $\Delta_\lambda^\beta$ is a correction to the RTA
from an iterative solution of the PBTE. 
Full details of the theory are described in Ref. \cite{ShengBTE}.
Diagonal elements of $\kappa$ along other directions are obtained by a unitary transformation (rotation) 
of the thermal conductivity tensor.

The calculation of the three-phonon matrix elements, 
needed for the calculation of $\tau_\lambda^0$ and $\Delta_\lambda^\beta$ \cite{ShengBTE},
requires the third-order (anharmonic) IFCs.
A 2$\times$2$\times$1 supercell is used to calculate
the anharmonic IFCs, which generates 3472 and 520 
atomic structure configurations for \ta and \zr, respectively.
Atomic interactions up to fifth-nearest neighbor are considered. 
Both the RTA and full iterative approach, are used to solve the phonon BTE, 
as implemented in the ShengBTE package \cite{ShengBTE, ShengBTE2}.
In the iterative approach, 
$\kappa$ is converged to a precision of 10$^{-5}$ between iterative steps. 
Convergence with respect to the k-point grid is also checked.
The converged Monkhorst-Pack grids are 14$\times$14$\times$6 for \ta and 8$\times$14$\times$8 for \zr.

\section{Results and discussions}
\subsection{Phonon dispersion}
\label{sec:phonon_dispersion}
The phonon dispersions of \ta and \zr are shown in Fig. \ref{fig:TaSe3_NbS3_unitcell}(e,f).
The color scheme indicates the absolute group velocity at each phonon \textbf{q}-vector and phonon branch
with the magnitudes given by the color bars.
The large number of atoms in the unit cells result in many low-frequency optical modes.
A number of the optical modes are highly dispersive.
Experimentally, the many optical modes
create complex Raman spectra with closely spaced peaks.
The red circles indicate the peaks from experimental Raman data.\cite{1982_Levy_raman,Zwick_1980}
Most of the experimental Raman peaks match well with the DFT calculated phonon 
frequencies at $\Gamma$ as shown in Fig. \ref{fig:TaSe3_NbS3_unitcell}(e,f).
The experimental and calculated values are listed in the Supplementary Information along with
images of the displacements and symmetry of each mode at $\Gamma$. 
Below, we first discuss the anisotropy of the acoustic modes, and then we discuss the nature of several
of the lower frequency, dispersive optical modes.

The anisotropy of the transition metal trichalcogenides has been a topic of long-term interest.
The degree with which a crystal behaves as quasi-1D or 2D depends on the strength of the interchain coupling,
and the interchain coupling manifests itself in the directional dispersions of both the electrons and phonons.
Larger coupling results in more dispersive bands and higher velocities.
Thus, the phonon velocities give one measure of the anisotropy of the crystals.
The velocities of the three acoustic modes along four high symmetry lines of the
Brillouin zone and the exact cross-chain direction are given in Table \ref{tab:velocities}.

For these crystals, the LA velocities along the chain directions ($\Gamma-Y$)
are high, and for TaSe$_3$, they are the maximum velocities
among all modes and all directions, as one would expect from a quasi-1D crystal structure. 
%
%
For \zr, the highest velocity phonon is the LA mode in the {\em cross-chain} direction.
This is consistent with its electronic anisotropy in which the highest Fermi velocities occur
for the Te p$_x$ bands in the cross-chain direction (electronic structure plots are shown in the Appendix).
%
%

The maximum velocities of the TA shear modes are highest along the chain directions for these crystals,
with the one exception of the TA$_2$ mode of \ta which has its highest velocity in the cross-chain direction.
Also, there is more nonlinearity to the TA mode dispersions, 
such that the maximum velocities of the TA modes occur at a finite frequency on the order of 1 THz.

Finally, the lowest velocity LA modes for both crystals are in the cross-plane direction.
For \ta, the lowest velocity TA modes are also in the cross-plane direction, 
and
for \zr, the lowest velocity TA modes are in the cross-chain direction.

The acoustic phonon velocities exhibit different degrees of anisotropy
for these two crystals.
%
%
If we consider, for example, the LA mode, then its anisotropy in \ta is significantly
larger than in \zr.
For \ta, 
the ratios of the maximum velocities in the chain ($v_y$), cross-chain ($v_{cc}$), 
and cross-plane ($v_{cp}$) directions are
$v_y / v_{cc} = 1.2$, 
$v_y / v_{cp} = 2.1$, and 
$v_{cc} / v_{cp} = 1.8$.
For \zr, the ratios are
$v_{cc} / v_y = 1.09$, 
$v_{cc} / v_{cp} = 1.8$, and
$v_y / v_{cp} = 1.3$.
Every ratio in \ta is greater than or equal to the corresponding one in \zr.
The LA mode of both materials 
appears to be more quasi-two-dimensional rather than quasi-one-dimensional,
since the in-plane anisotropy, as characterized by the ratio $v_y/v_{cc}$ or $v_{cc}/v_{y}$, 
is considerably less than the 
cross-plane anisotropy as characterized by the ratios $v_y / v_{cp}$ and $v_{cc} / v_{cp}$.
The anisotropy of the maximum velocities of the TA modes is always less than that of the LA modes.
For example, in \ta, the TA$_1$ velocity ratios are
$v_y / v_{cc} = 1.2$, 
$v_y / v_{cp} = 1.5$, and 
$v_{cc} / v_{cp} = 1.2$;
and
the TA$_2$ ratios are
$v_{cc} / v_y = 1.03$, 
$v_{cc} / v_{cp} = 1.3$, and
$v_y / v_{cp} = 1.2$.

We now consider the optical modes.
The vibrational modes of \ta, shown in Fig. \ref{fig:TaSe3_NbS3_unitcell}(e), can be represented at $\Gamma$ as
\cite{1982_Levy_raman}, 
\begin{equation}
\Gamma_\mathrm{TaSe_3} = 8 A_u + 8 B_g + 16 B_u + 16 A_g.
\end{equation}
The $8A_u + 8B_g$ modes have vibrations along the chain axis ($b$ axis),  
and the $16B_u + 16A_g$ modes are polarized in the $\av-\cv$ plane.
Illustrations of the displacements of the modes based on the point group symmetry of the isolated chains
are shown in Ref. [\onlinecite{Wieting_in_Qu_1D_Conds79}].
For each mode at $\Gamma$, the frequency, symmetry and
images of the displacements calculated for the periodic unit cell
are provided in the Supplementary Information (SI).
The measured Raman frequency taken from the literature is also listed for each $A_g$ 
and $B_g$ mode.

Relevant to thermal transport,
there are several low-frequency optical modes that have relatively large dispersion
along the chain ($\Gamma-Y$) direction with maximum velocities in the range of $1-1.5$ km/s. 
The three low-frequency optical modes with the largest dispersions (highest velocities)
are $B_g$ or $A_u$ modes with displacements along the chain direction.
Modes 5 (56.3 \cmu) ($B_g$), 6 (57.1 \cmu) ($A_u$), and 8 (74.5 \cmu) ($B_g$)
at $\Gamma$ (counting upwards from 0 frequency) are examples of such modes.
Mode 4 (52.0 \cmu, $A_g$), with displacements in the $a-c$ plane is slightly 
less dispersive immediately near $\Gamma$, but it has a relatively constant 
velocity of $\sim 1$ km/s along the entire $\Gamma-Y$ line.
Above 75 \cmu, the longitudinal acoustic mode appears to hybridize with many optical modes 
resulting in significant dispersion along $\Gamma-Y$ for modes between 75 \cmu and 130 \cmu.

The other types of optical modes are the $A_g$ and $B_u$ modes with either
rotation or libration type 
non-collinear displacements in the $a-c$ plane perpendicular to the 1D chains.
The 4th (51.7 \cmu) ($A_g$), 7th (70.2 \cmu) ($A_u$), and 9th (82.7 \cmu) ($A_g$) 
modes at $\Gamma$ correspond to such modes. 
In mode 4, each half of the unit cell, on either side of the van der Waals gap, 
has a rotational type displacement,
and the rotational displacements of each half are out of phase by $180^\circ$.
This is typical of all of the rotational type $A$ modes with displacements on the ${\bf a}-{\bf c}$ plane.

Fig. \ref{fig:TaSe3_NbS3_unitcell}(f) shows the phonon dispersion of \zr.
For \zr, there are 24 vibrational modes which can be represented at $\Gamma$ as \cite{Zwick_1980}
\begin{equation}
\Gamma_\mathrm{ZrTe_3} = 4 A_u + 4 B_g + 8 B_u + 8 A_g.
\end{equation}
%
%
Similar to \ta, the displacements of the $4A_u + 4B_g$ 
modes have vibrations along the chain ($b$) axis,
and the $8B_u + 8A_g$ modes have displacements on the $a-c$ plane.
There are 4 optical modes with relatively large dispersions along the chain ($\Gamma-Y$):
modes 5 (64.0 \cmu, $B_g$), 8 (69.1 \cmu, $B_g$), 9 (70.4 \cmu, $A_g$), and 12 (91.1 \cmu, $A_g$). 
For \zr, both types of displacements give rise to 
these relatively high velocity optical modes. 
Modes 5 and 8 have displacements along the chain and modes 9 and 12 have displacements on
the $a-c$ plane.
Between 75 \cmu and 110 \cmu, hybridization occurs between the longitudinal acoustic
mode and the optical modes resulting in significant dispersion of the modes in that frequency
window.

Although the highest velocity acoustic mode is in the cross-chain ($\Gamma-X$)
direction, there are fewer dispersive optical modes along $\Gamma-X$ than in the chain direction.
%
%
For this reason, the thermal conductivity is highest in the chain direction, since, as
we shall see, a significant proportion of the heat is carried by the optical modes.

\begin{figure}[]
\centering \includegraphics[width=3.5in]{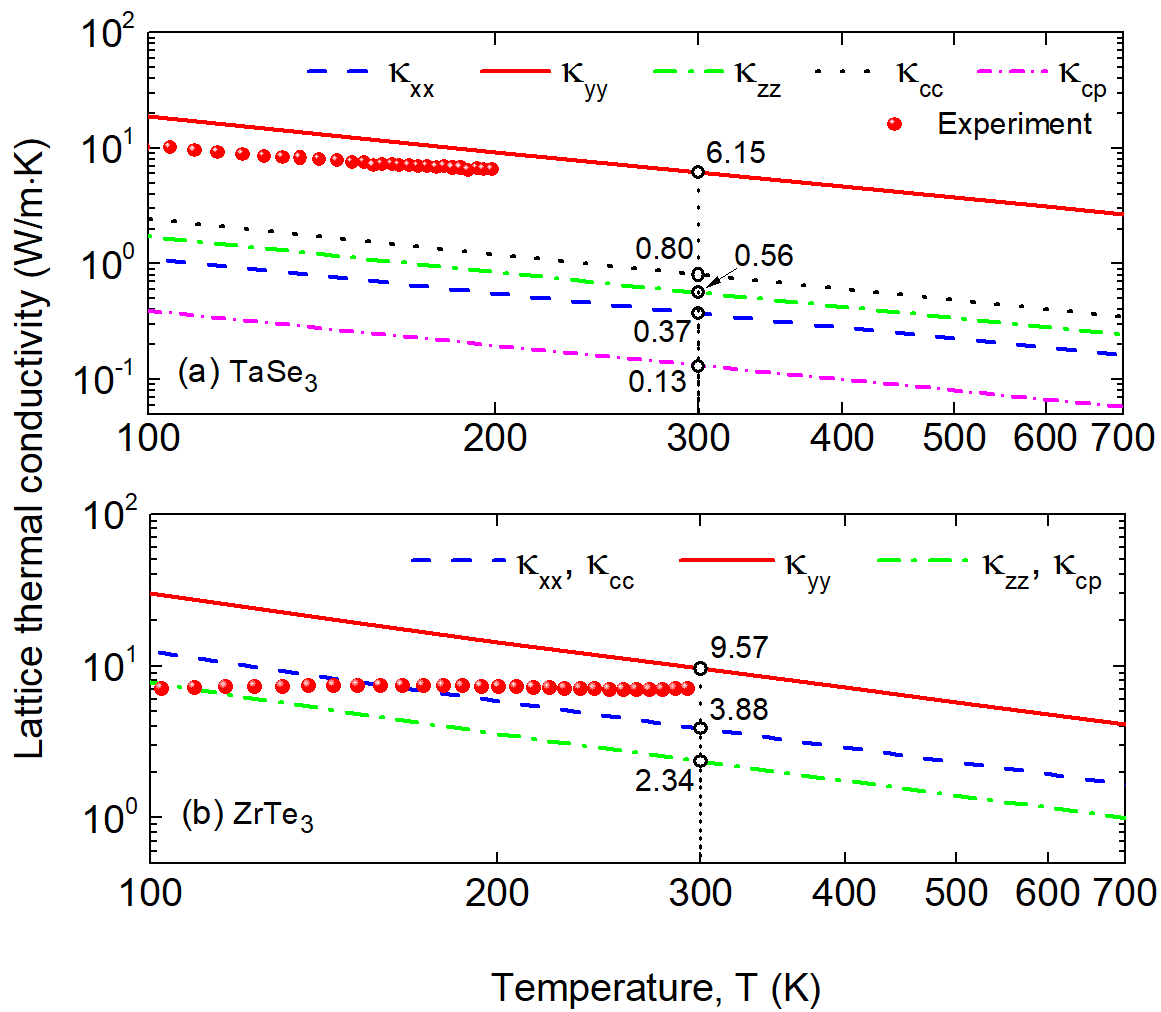}
\caption{
The three diagonal components of the lattice thermal conductivity 
tensor versus temperature for (a) \ta and (b) \zr.
For \ta, the diagonal components of the transformed thermal 
conductivity tensor in the cross-chain ($\kappa_{cc}$) and
cross-plane ($\kappa_{cp}$) directions are also shown.
For \zr, $\kappa_{cc} \equiv \kappa_{xx}$ and $\kappa_{cp} \equiv \kappa_{zz}$.
The various components are indicated by the legends.
The values for $T=300$ K are labeled on the plots in units of Wm$^{-1}$K$^{-1}$.
The red circles are experimental thermal conductivity values for (a) \ta and (b) 
polycrystalline \zr taken from the 
literature \cite{2010_thermal_conductivity,Hooda_2019}.
}
\label{fig:TC}
\end{figure}

\begin{table*}[!ht]
\centering
\small
\caption{
Thermal conductivities from this paper and from the literature. 
} 
\label{Table:comparison}
\begin{tabularx}{1\textwidth}{ccccc}
\hline
\hline
\multicolumn{1}{>{\centering\arraybackslash}m{17.4mm}}{{Type}}
	& \multicolumn{1}{>{\centering\arraybackslash}m{20mm}}{{Materials}} 
    & \multicolumn{1}{>{\centering\arraybackslash}m{33.5mm}}{{Thermal Conductivity (Wm$^{-1}$K$^{-1}$)}} 
    & \multicolumn{1}{>{\centering\arraybackslash}m{60mm}}{{Remarks}}
    & \multicolumn{1}{>{\centering\arraybackslash}m{33mm}}{{Reference}}\\
    \hline

\hline
\mbox{ } & \ta & 6.2, 0.80, 0.13 & Chain, cross-chain and cross-plane directions &  This paper \\
\mbox{ } & \zr & 9.6, 3.9, 2.3 & Chain, cross-chain and cross-plane directions &  This paper \\
\mbox{ } & \ta & 4.8 & Experimental result (extrapolated to 300K) &  \cite{2010_thermal_conductivity} \\
\mbox{ } & NbSe$_3$ & 24 & Bulk Material & \cite{BRILL_1981} \\
Quasi-1D & TiS$_3$  & 5.8, 2.8 & Chain and cross-chain & \cite{TiS3_Heat_OpPhonons_NLett20} \\
\mbox{ } & ZrTe$_3$ & 7 &  Bulk polycrystalline & \cite{Hooda_2019} \\
\mbox{ } & ZrTe$_5$ & 11.2 & Bulk material &  \cite{Smontara_1985}  \\
\mbox{ } & ZrTe$_5$ & 3.9, 1.9, 0.4 & Chain, cross-chain and cross-plane directions &  \cite{Zhu_2018} \\
\vspace{0.2cm}
\mbox{ } & ZrTe$_5$ & 2.2 & Bulk polycrystalline &  \cite{Hooda_2017}\\
\mbox{ } & HfS$_2$  & 9 & First principle study & \cite{Yumnam_2015} \\
\mbox{ } & HfSe$_2$ & 8 & First principle study & \cite{Yumnam_2015} \\
\mbox{ } & MoS$_2$  & 1 & Pristine & \cite{kim_2010} \\
\mbox{ } & MoSe$_2$ & 0.9 & Pristine &  \cite{kim_2010} \\
\mbox{ } & MoSe$_2$ & 2.3 & Compacted polycrystalline & \cite{Brixner_1962} \\
\mbox{ } & MoTe$_2$ & 1.9 & Compacted polycrystalline & \cite{Brixner_1962} \\
\mbox{ } & NbS$_2$  & 11.8 & Bulk material & \cite{nishio_1994}\\
\mbox{ } & NbSe$_2$ & 2.1 & Compacted polycrystalline & \cite{Brixner_1962} \\
\mbox{ } & NbTe$_2$ & 1.9 & Compacted polycrystalline & \cite{Brixner_1962} \\
\mbox{ } & TaS$_2$  & 5  & Bulk material &  \cite{Regueiro_1985} \\
2D		 & TaSe$_2$ & 16 & Bulk material & . \cite{Regueiro_1985,Yan_2013} \\
\mbox{ } & TaSe$_2$ & 1.7 & Compacted polycrystalline&  \cite{Brixner_1962} \\
\mbox{ } & TaTe$_2$ & 1.4 & Compacted polycrystalline & \cite{Brixner_1962} \\
\mbox{ } & TiS$_2$  & 6.8, 4.2 & In-plane and out-of-plane & \cite{Imai_2001} \\
\mbox{ } & WSe$_2$  & 1.7 & Compacted polycrystalline &  \cite{Brixner_1962} \\
\mbox{ } & WS$_2$   & 2.2 & Pristine & \cite{kim_2010} \\
\mbox{ } & WSe$_2$  & 0.8 & Pristine & \cite{kim_2010} \\
\mbox{ } & WTe$_2$  & 1.6 & Compacted polycrystalline & \cite{Brixner_1962}\\
\mbox{ } & ZrSe$_2$ & 10 & First principle study &  \cite{Yumnam_2015} \\
\mbox{ } & ZrS$_2$  & 18 & First principle study &  \cite{Yumnam_2015} \\
\hline
\end{tabularx}
\end{table*}

\subsection{Thermal conductivity}
Figure \ref{fig:TC} shows the lattice thermal 
conductivity for \ta and \zr calculated from the PBTE using the full iterative approach.
The iterative approach gives values slightly higher than those from
the relaxation time approximation, and a comparison of the results from the two approaches
is given in the Appendix.
Below, we discuss the magnitudes and anisotropies of the thermal conductivities using
the $T=300$ K values shown in Fig. \ref{fig:TC}.
The lattice thermal conductivities in the chain directions ($\kappa_{yy}$) of \ta and \zr,
are 6.15 Wm$^{-1}$K$^{-1}$ and 9.57 Wm$^{-1}$K$^{-1}$, respectively. 
These values are similar to those 
of other metallic 2D TMD materials as shown in Table \ref{Table:comparison}.
The calculated lattice thermal conductivities of \ta and \zr follow a T$^{-1}$ dependence expected
when the thermal conductivity is limited by three-phonon scattering. \cite{Yao_2017}
The diagonal elements of $\kappa$ are fitted to the function $c_1 + c_2 T^{-1}$, and
the values of the coefficients are tabulated in the Appendix.
The root mean square errors of all fits to the numerical data are less than 
$3 \times 10^{-15}$ Wm$^{-1}$K$^{-1}$.
The anisotropy of the thermal conductivity in \ta, 
as indicated by the ratios 
$\ky/\kcc = 7.7$ and $\ky / \kcp = 47$ is large.
For \zr, the anisotropy given by the ratios 
$\ky/\kcc = 2.5$ and $\ky / \kcp = 4.1$ is considerably less.
Unlike the electrical conductivity which is maximum in the cross-chain direction,
the thermal conductivity of \zr is maximum in the chain direction,
even though the maximum phonon velocity occurs at very low frequencies 
for the LA mode in the cross-chain direction.
The experimental reports of thermal conductivity of 
quasi-1D materials are rare, due to the
difficulty in measuring the thermal conduction in 
such ribbon-like geometries \cite{2010_thermal_conductivity}.
To our knowledge, only one experimental study of the thermal conductivity
of \ta has been reported using a parallel thermal
conductance (PTC) technique \cite{2001_PTC},
and the data points are shown in Fig. \ref{fig:TC}(a)\cite{2010_thermal_conductivity}.
The experimental values are lower than the calculated values, and the comparison
improves at higher temperatures when the phonon-phonon
scattering mechanism starts to dominate.
At T = 200 K, the experimental and calculated values are
6.62 Wm$^{-1}$K$^{-1}$ and 9.11 Wm$^{-1}$K$^{-1}$, respectively.
From the experimental side, the accuracy of 
PTC technique to handle chain-structures like \ta 
is under question \cite{2010_thermal_conductivity}. 
Missing from the calculations are the effects of
impurities, surface scattering, finite grain sizes, and random isotopes \cite{lindsay2013phonon}. 
For \zr, there is one reported measurement of the thermal conductivity
in a bulk, polycrystalline sample with a value of 7 \wmk at room temperature which is 
similar to our calculated value of 9 \wmk. 

\begin{figure}[ht]
\centering \includegraphics[width=3.5in]{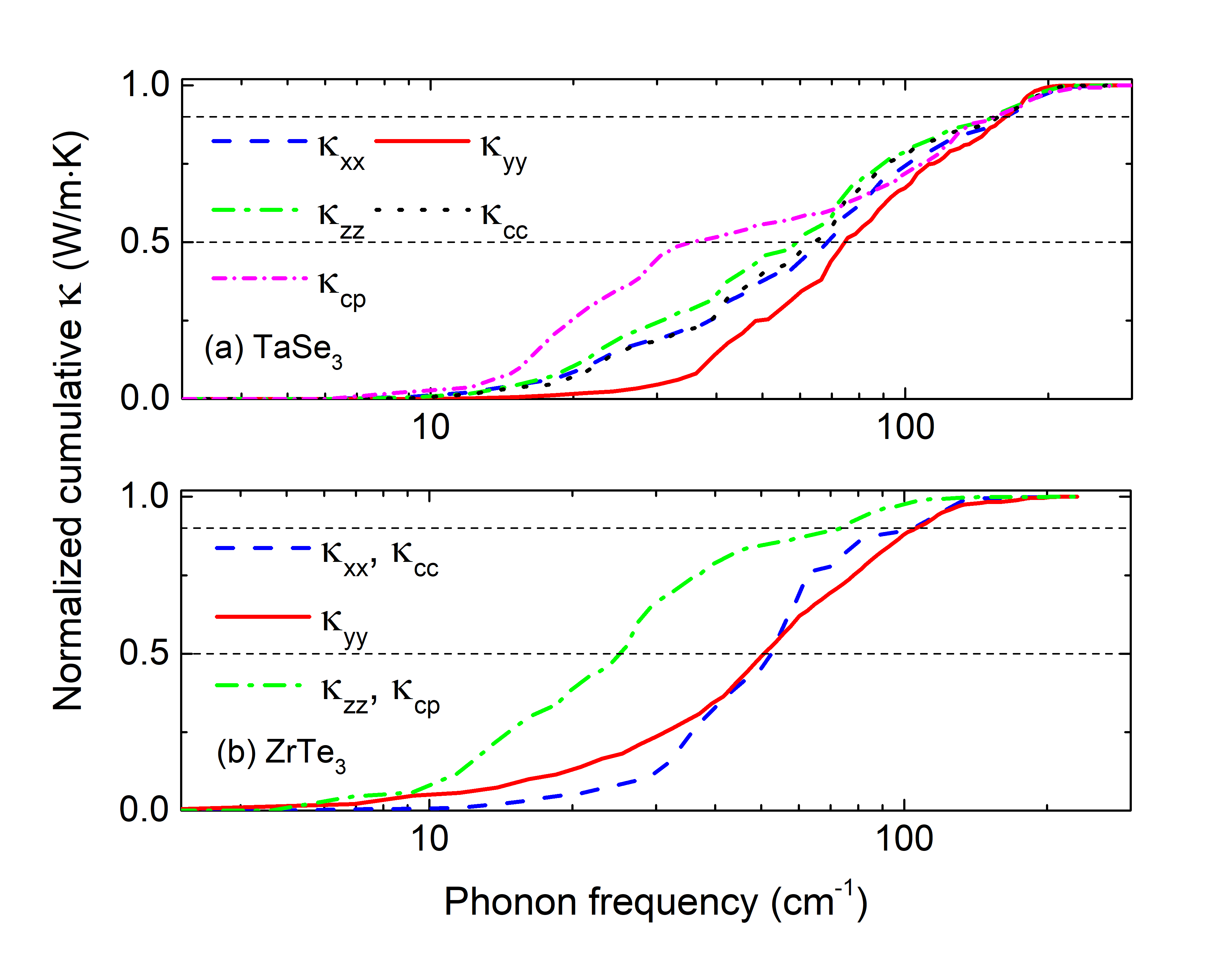}
\caption{
Thermal conductivity values at $T=300$ K, normalized to their maximum values 
along various crystallographic directions, 
as cumulative functions of the phonon frequency for (a) TaSe$_3$ and (b) ZrTe$_3$.
For ZrTe$_3$, the values for the cross-chain (cc) and cross-plane (cp) directions
are indistinguishable from the values for $\kappa_{xx}$ and $\kappa_{zz}$, respectively. 
Reference horizontal lines are at 0.5 and 0.9.
}
\label{fig:Cumulative_Kappa}
\end{figure}

\begin{figure}[ht]
\centering \includegraphics[width=3.6in]{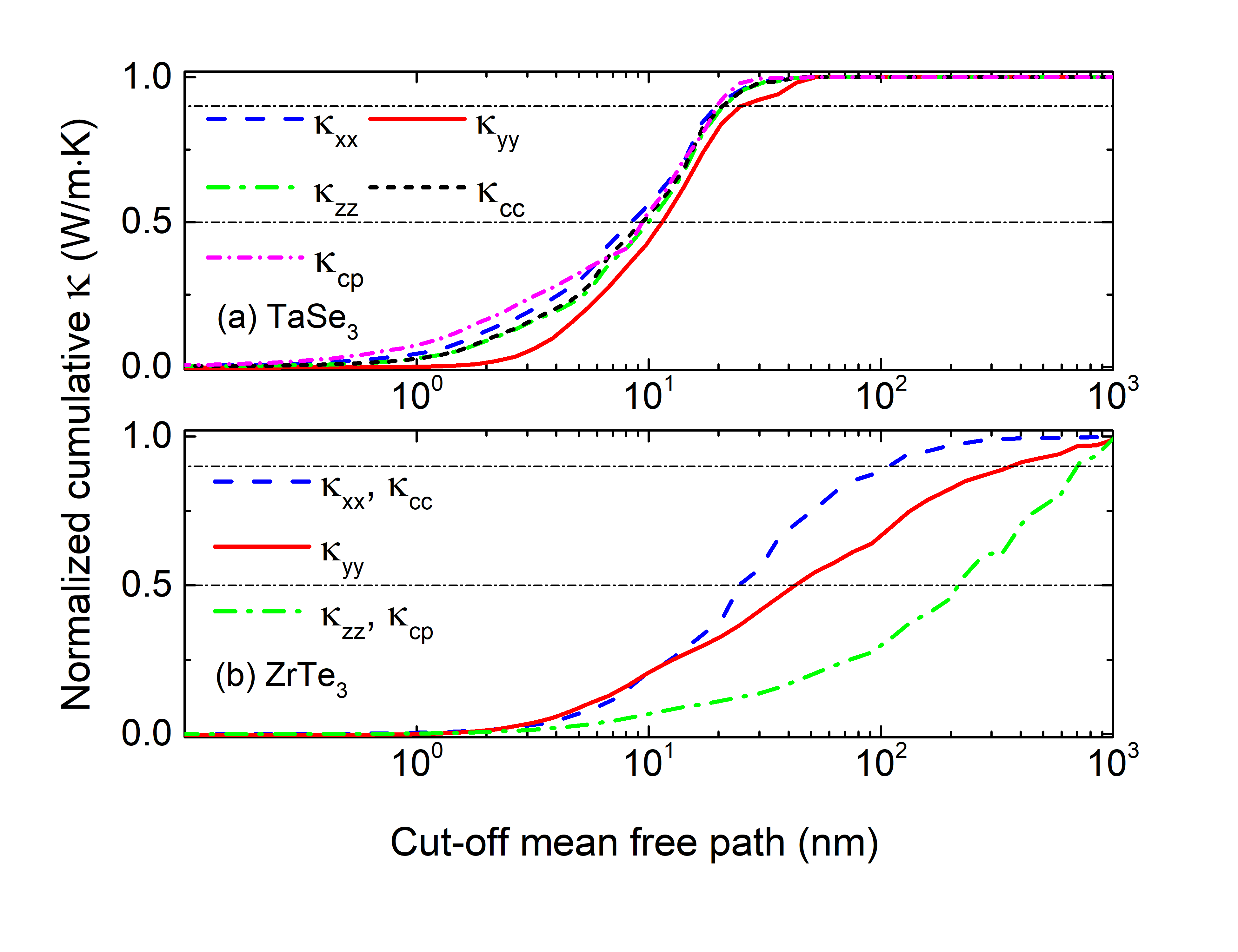}
\caption{
Thermal conductivity values at $T=300$ K, normalized to its maximum value 
along various crystallographic directions, 
as functions of the cut-off mean free path for (a) TaSe$_3$ and (b) ZrTe$_3$.
Reference horizontal lines are at 0.5 and 0.9.
}
\label{fig:MFP}
\end{figure}
\begin{table}[b]
\centering
\def\arraystretch{1.2}
\caption{Representative mean free path 
($\lambda_\text{R}$) of heat carrying phonons 
in \ta and \zr,  at T = 300 K
in the 3 directions indicated in the header.} \label{Table:MFP}
\begin{tabularx}{0.475\textwidth}{XXXXXX}
\hline
\hline
& \multicolumn{5}{c}{$\lambda_\text{R}$ (nm)} \\
	 \cline{2-6}
& x & y & z & cross-chain & cross-plane\\
\hline
\ta & 7.3 & 10.6 &  8.5 & 8.2 & 7.1\\
\zr & 25.3 & 43.7 &  178 & 25.3 & 178\\
\hline
\hline
\end{tabularx}
\end{table}

\begin{figure*}[ht]
\centering \includegraphics[height = 3.0in]{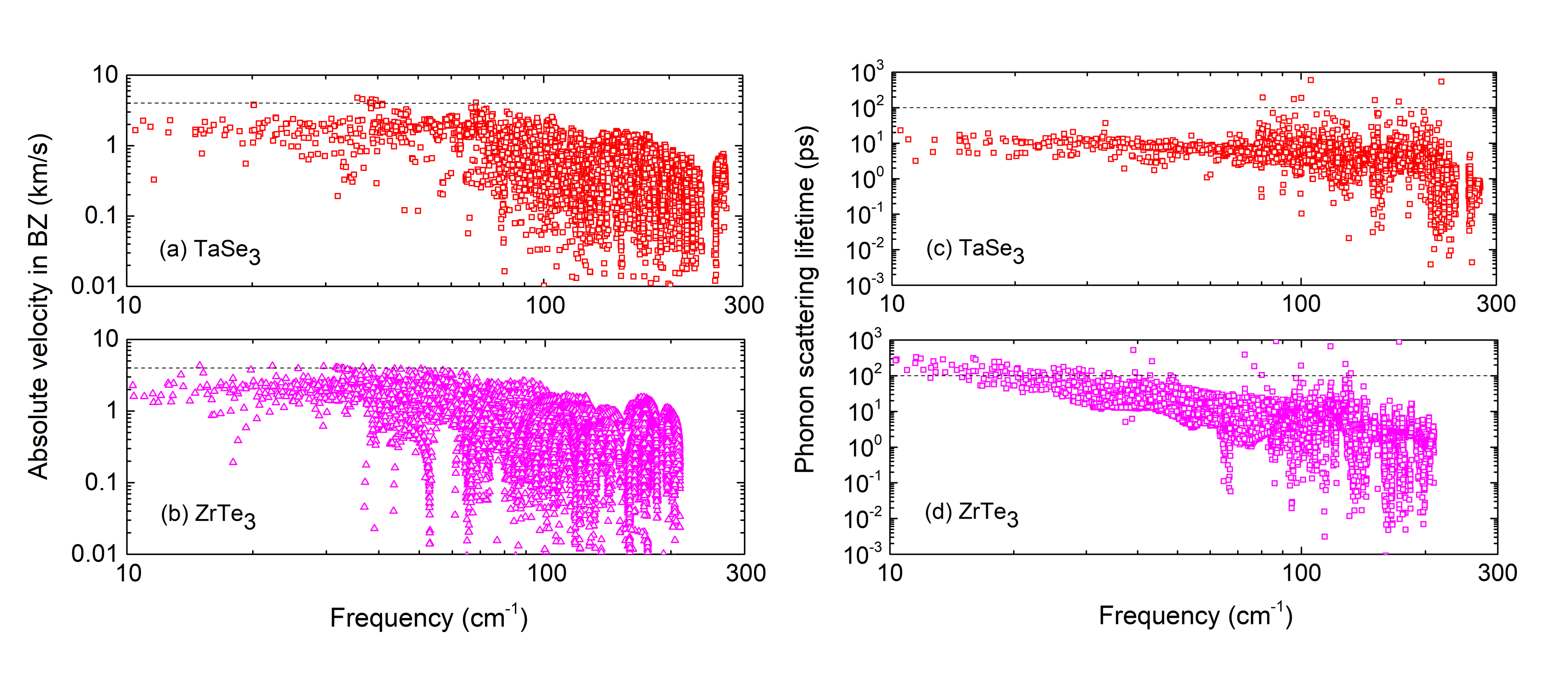}
\caption{Absolute velocity of phonon 
modes in the irreducible Brillouin zone of (a) \ta and (b) \zr.
The reference horizontal lines are at 4 km/s.
Scattering lifetimes of the phonon modes at $T=300$ K for (c) \ta and (d) \zr. The reference
horizontal lines are at 100 ps. 
}
\label{fig:lifetime_vBZ}
\end{figure*}

Order-of-magnitude estimates of the electronic components of the 
thermal conductivities can be obtained from the Wiedemann-Franz law 
using the Sommerfeld value of the Lorenz number, 2.44 $\times$ 10$^{-8}$ W$\Omega$ K$^{-2}$.
Values for the Lorenz number do vary, but for room temperature metals, they tend to lie within a
range of 0.6 to 2 times the Sommerfeld value. \cite{Lorenz_no_review93}
%
%
Using the values for the electrical conductivities from Sec. \ref{sec:crystal_structures}, 
the estimated room-temperature electrical component of the thermal conductivity along the chain
direction for \ta is 1.2 Wm$^{-1}$K$^{-1}$.
For \zr, the estimated room-temperature electrical component of the thermal conductivity along the chain
direction ranges from 2.9 to 5.1 Wm$^{-1}$K$^{-1}$, and along the cross-chain direction, it ranges from
4.1 to 9.6 Wm$^{-1}$K$^{-1}$.
For comparison, the corresponding values of the room temperature lattice thermal conductivities for \ta and \zr
along the chain direction are 6.2 and 9.6 Wm$^{-1}$K$^{-1}$, respectively, 
and for \zr along the cross chain direction, it is 3.9  Wm$^{-1}$K$^{-1}$.
Thus, the Wiedemann-Franz law with the Sommerfeld approximation for the Lorenz number
gives room-temperature electrical components of the thermal conductivities that are on the same order as
the lattice components.

To understand the contribution of the phonon modes and frequencies to the 
lattice thermal conductivity, we consider the room-temperature, normalized 
thermal conductivity as a cumulative function of phonon frequency in Fig. \ref{fig:Cumulative_Kappa}. 
In \ta, 50\% of the thermal conductivity in the chain direction ($\kappa_{yy}$)
is contributed by modes with frequencies below 75 cm$^{-1}$. 
%
The thermal conductivity in the $a-c$ plane
has a larger contribution from lower frequency phonons.
In the cross-chain direction,
50\% of $\kappa_{cc}$ is contributed by modes with frequencies below 64 \cmu,
and in the cross-plane direction,
50\% of $\kappa_{cp}$ is contributed by modes with frequencies below 36 \cmu.

In \zr, the heat is carried by lower frequency phonons than in \ta.
In \zr, 50\% of thermal conductivity along both the chain direction ($\kappa_{yy}$)
and cross-chain direction ($\kappa_{cc}$)
is contributed by phonons with frequencies below 52 \cmu.
%
In the cross-plane direction 50\% of $\kappa_{cp}$ is contributed by modes with
frequencies below 26 \cmu. 

For both of these materials, 
a significant percentage of the heat at room temperature is carried by the optical modes.
This phenomenon has also recently been observed in TiS$_3$ \cite{TiS3_Heat_OpPhonons_NLett20}. 
Consider the chain direction, $\Gamma-Y$. 
In \ta, essentially all of the modes above 75 \cmu
are optical modes, as can be seen in Fig. \ref{fig:TaSe3_NbS3_unitcell}(e). 
Thus, approximately 50\% of the heat in \ta at room temperature is carried by optical modes.
In \zr, 
the spectrum along the chain direction above
64 \cmu is composed primarily of optical modes, and it carries 33\% of the heat.
%
%

%
Another way to analyze the physics of the heat current carried by phonons is
to calculate the normalized $\kappa$ as a function of the cut-off mean free path. 
The results are shown in Fig. \ref{fig:MFP}.
For \ta, 90\% of the thermal conductivity in 
the chain direction is carried by phonons with a mean free path of 25 nm or less.
In both the cross-chain and cross-plane
directions 90\% of the thermal conductivity
is carried by phonons with mean free paths of 21 nm or less.
The corresponding cut-off mean free paths for \zr are approximately an order of magnitude greater.
For \zr, in the chain, cross-chain, and cross-plane directions,
90\% of the heat is carried by phonons with mean free paths less than or equal to
365 nm, 110 nm, and 700 nm, respectively.

To obtain a representative mean free path ($\lambda_\text{R}$) of 
heat-carrying phonons, the cumulative $\kappa$ with
respect to the cut-off mean free path ($\lambda_\text{max}$) in Fig. \ref{fig:MFP},
is fitted to 
a single parametric function, \cite{ShengBTE,wang_2017} 
\begin{equation}
\kappa\,(\lambda_\text{max}) = \frac{\kappa_0}{1+\lambda_\text{R}/\lambda_\text{max}},
\end{equation}
where, $\kappa_0$ is the maximum thermal conductivity.
The values for $\lambda_R$ for the different transport directions are
tabulated in Table \ref{Table:MFP}. 
They correspond closely to the cut off mean free paths in Fig. \ref{fig:MFP} that account
for 50\% of the thermal conductivity in each direction.
These values serve as indicators as to how and when the thermal conductivity will be affected by 
geometrical scaling.
When dimensions are reduced below $\lambda_R$, the thermal conductivity will be reduced.
The representative mean free path ($\lambda_R$) for \ta is lower than that of \zr,
as shown in Table \ref{Table:MFP}. 
This is consistent with the fact that, in the low frequency ($0 \sim 50$ cm$^{-1}$) region, 
the phonon scattering lifetimes in \ta are approximately one order of magnitude shorter than those of \zr,
as discussed below and shown in Fig. \ref{fig:lifetime_vBZ}(c,d).  
The cross-plane thermal conductivity of \zr has the longest representative mean free path of 178 cm$^{-1}$.
This is consistent with the fact that the phonons contributing to the cross-plane 
thermal conductivity are very low frequency (50\% of the heat is carried by phonons with
frequencies below 25 \cmu). 
This frequency range is well below the optical branches, so that three phonon scattering is strongly
restricted by energy and momentum conservation.

To gain further insight into why there is a significant difference between
the thermal conductivity in these materials, we inspect the 
phonon velocity as well as the phonon lifetimes
of the thermal modes.
Figure \ref{fig:lifetime_vBZ}(a,b) shows the 
absolute velocity distributions of the
heat-carrying phonons, for each phonon mode 
inside the irreducible Brillouin zone (BZ),
and the reference horizontal lines are at 4 km/s.
The lifetimes are shown in Fig. \ref{fig:lifetime_vBZ}(c,d),
and the reference horizontal lines are at 100 ps.
In the low frequency range $\lesssim 50$ \cmu, \zr has higher velocity 
phonons with longer lifetimes than those of \ta.

In this low frequency range, the phonon lifetimes of \zr
are approximately one order of magnitude longer than those of \ta.
For \zr, the optical modes are close to or above 50 \cmu.
For \ta, 
there are several optical branches that fall below 50 \cmu accompanied by a number of
band crossings and anti-crossings. 
Also, since the unit cell of \ta is twice as large in the $a-c$ plane as that of \zr,
there are features in the acoustic dispersion along $\Gamma - X$ and $\Gamma - Z$ that resemble
zone-folding in which the acoustic branches fold back at the zone boundary and return to
$\Gamma$.
Both of these features result in a greater number of channels for low-frequency phonon 
relaxation in \ta compared to those in \zr and correspondingly lower phonon lifetimes.
A simple illustration of how zone-folding opens new phonon relaxation channels is shown in 
Ref. [\onlinecite{li2014thermal}].

\section{Summary and Conclusion}
The phonon dispersions and lattice thermal 
conductivities of \ta and \zr are determined using density functional theory and the phonon BTE. 
The anisotropy of the LA acoustic phonons, as characterized by the ratios of the
chain, cross-chain, and cross-plane velocities, is considerably larger
in \ta than in \zr.
The anisotropy of the maximum velocities of the TA modes is always less.
The maximum LA velocity in \zr occurs in the cross-chain direction, and this is consistent with the
strong cross-chain bonding that gives rise to large Fermi velocities.

However, the thermal conductivity for both crystals is maximum in the chain direction.
The thermal conductivity of \zr is larger than that of \ta in each of the
three directions: chain, cross-chain, and cross-plane;
and it is considerably more isotropic.
For \ta (\zr), the room temperature diagonal thermal conductivity values in the three directions are
$\ky$ = 6.2 (9.6), $\kcc$ = 0.80 (3.9), and $\kcp$ = 0.13 (2.3) Wm$^{-1}$K$^{-1}$.

A significant percentage of the heat at room temperature is carried by the optical phonons.
In \ta, the spectrum along the chain direction above 75 \cmu is composed of the optical branches 
and this part of the spectrum carries 50\% of the heat at room temperature.
In \zr, the spectrum along the chain direction above 64 \cmu is composed primarily of optical branches, and it
carries 33\% of the heat at room temperature.
For \ta (\zr) along the chain direction at $T=300$ K, 
50\% of the heat is carried by phonons below 75 (50) \cmu, and this part of the phonon spectra
consists primarily of the acoustic branches.

The differences between the two materials in their phonon velocities and lifetimes are most apparent in
the low frequency range $\lesssim 50$ \cmu.
In this frequency range, the maximum phonon velocities of \zr are approximately a factor of
2 greater than those of \ta and the phonon lifetimes in \zr are approximately an order of magnitude greater
than those in \ta.
The longer lifetimes result in considerably longer mean free paths in \zr compared to those in \ta.
The representative mean free paths in the chain, cross-chain, and cross-plain directions for \ta (\zr) are
10.6 (43.7), 8.2 (25.3), and 7.1 (178) nm, respectively. 
The shorter lifetimes in the low frequency range of \ta are consistent with the presence of optical branches and zone-folding
features of the acoustic branches that arise due to the doubling of the \ta unit cell in the $a-c$ plane compared to
the unit cell of \zr.
Both of these features serve to introduce more scattering channels for low frequency phonon relaxation.

\section{Acknowledgment}
This work was supported in part by the NSF under Grant No EFRI-1433395.
Calculations of ZrTe$_3$ were supported, in part, by 
Spins and Heat in Nanoscale Electronic Systems (SHINES) an 
Energy Frontier Research Center funded by the U.S. Department of Energy, Office of Science, 
Basic Energy Sciences under Award No DE-SC0012670. 
This work used the Extreme Science and Engineering Discovery
Environment (XSEDE)\cite{towns2014xsede}, which is supported by National
Science Foundation Grant No. ACI-1548562 and Allocation
ID TG-DMR130081.

\setcounter{figure}{5}
\setcounter{table}{3}

\section*{Appendix: ADDITIONAL TABLES AND FIGURES}
\label{sec:appendix}
The calculated and experimental lattice constants and angles for \ta and \zr 
are provided in Table \ref{tab:lattice}.
The room temperature values of the diagonal elements of the thermal conductivity tensors 
calculated from the RTA and the full iterative approaches are listed in Table \ref{tab:RTA},
and the comparison for all temperatures is shown in Fig. \ref{fig:RTA}.
The fitting coefficients for the temperature dependent thermal conductivities
$\kappa(T) = c_1 + c_2 T^{-1}$ are provided in Table \ref{tab:fits},
and the plot showing the quality of the fits is given in Fig. \ref{fig:fitted}.
Electronic structure plots for \ta and \zr are shown in Fig. \ref{fig:Ek}.
The convergence of the lattice thermal conductivities for different $q$-point grids is shown
in Fig. \ref{fig:TC-conv}.
\begin{table*}[]
\centering
\def\arraystretch{1.5}
\caption{Lattice constants (\AA) and angles (degrees) for \ta and \zr.
The values labeled ``This paper'' are the values obtained after structure relaxation. The experimental values in the corresponding literature were obtained at room temperature.}
\vspace{0.1cm}
\label{tab:lattice}
\setlength{\tabcolsep}{18pt}
\begin{tabular}{c  c  c  c  c  c  c  c}
\hline
\hline
Crystal & Remark & $a$ & $b$ & $c$ & $\alpha$ & $\beta$ & $\gamma$\\
\hline
\ta & This paper & 10.452 & 3.508 & 9.875 & 90 & 106.36 & 90\\
\mbox{ } & Experiment \cite{Bjerkelund_1964,1965_Crystal_structure} & 10.411  & 3.494 & 9.836  & 90 & 106.36  & 90\\
\zr & This paper & 5.915 &  3.882 & 10.152 & 90 & 97.94 & 90\\
\mbox{ } & Experiment \cite{1991_Furuseth} & 5.895  & 3.926 & 10.104  & 90 & 97.93  & 90\\
\hline
\end{tabular}
\end{table*}

\begin{table*}[]
\centering
\def\arraystretch{1.5}
\caption{Diagonal elements of the lattice thermal conductivity tensors 
calculated from the RTA and iterative methods at $T=300$ K.
}
\label{tab:RTA}
\setlength{\tabcolsep}{38pt}
\begin{tabular}{c  c  c  c  c}
\hline
\hline
 	& Method & $\kx$ & $\ky$ & $\kz$ \\
\hline

\ta 	 & RTA & 0.3718  & 4.0423  & 0.5644 \\
\mbox{ } & Iterative & 0.3722 & 6.1522 & 0.5643 \\
\hline
\zr 	 & RTA	& 3.6621	& 8.2807  &	2.2823 \\
\mbox{ } & Iterative	& 3.8764  & 9.5748  &	2.3435 \\
\hline
\end{tabular}
\end{table*}

\begin{table*}[]
\centering
\def\arraystretch{1.5}
\caption{Fitting coefficients $c_1$ (Wm$^{-1}$K$^{-1}$) and $c_2$ (Wm$^{-1}$) 
for the temperature dependence of 
the diagonal elements of $\kappa$ for \ta and \zr
given by the expression $\kappa = c_1 + c_2 T^{-1}$.
For \ta, the coefficients for $\kappa_{cc}$ and $\kappa_{cp}$ are also shown.
For \zr, these are the same as $\kappa_{xx}$ and $\kappa_{zz}$, respectively. 
}
\label{tab:fits}
\setlength{\tabcolsep}{20pt}
\begin{tabular}{ c  c  c  c  c  c  c}
\hline
 & coefficient & $\kx$ & $\ky$ & $\kz$ 	& $\kappa_{cc}$ & $\kappa_{cp}$ \\
\hline
\ta & $c_1$  & 0.003989 & -0.02828 & -0.003387 & -0.001137 & 0.001768 \\ 
    & $c_2$  & 110.4 & 1880. & 171.9 & 243.2 & 39.10 \\
    
\hline
\zr & $c_1$ & -0.1578 & -0.1457 & -0.1441 & - & - \\
    & $c_2$ & 1241 & 2956 & 772.3 & - & - \\
\hline
\end{tabular}
\end{table*}

\begin{figure*}[]
\centering \includegraphics[height=3.5in]{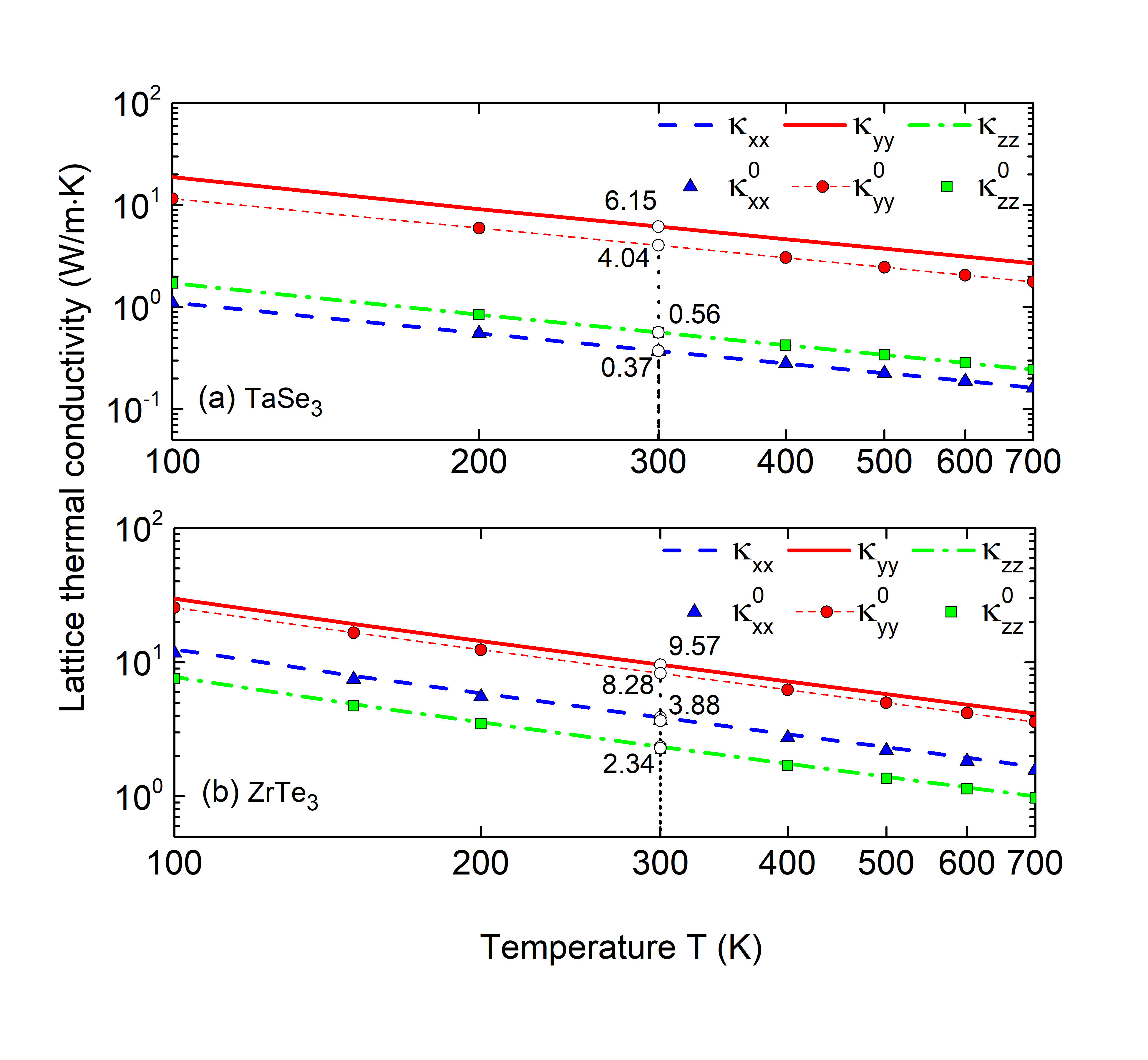}
\caption{
Lattice thermal conductivities of (a) \ta and (b) \zr calculated from the RTA and iterative approaches
as indicated by the legends. 
The curves labeled $\kappa^{0}$ are calculated using the RTA and the curves labeled $\kappa$ are
calculated from the full iterative method.
}
\label{fig:RTA}
\end{figure*}

\begin{figure*}[]
\centering \includegraphics[height=3.5in]{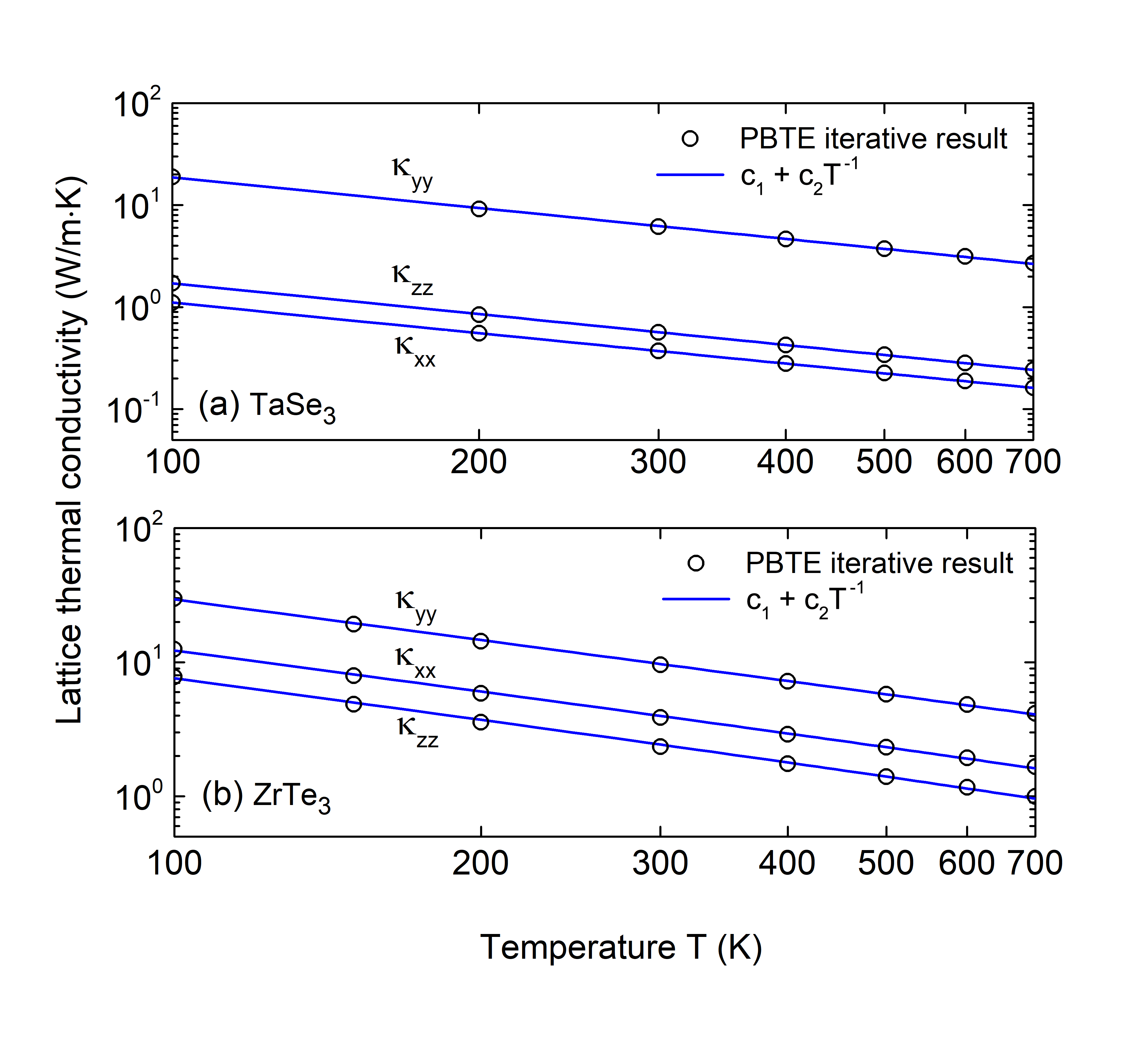}
\caption{
Fitted thermal conductivities for (a) \ta and (b) \zr using the coefficients 
listed in Table \ref{tab:fits}. 
The circles show the numerically calculated values, 
and the solid lines show the analytical fits. 
}
\label{fig:fitted}
\end{figure*}

\begin{figure*}[]
\centering \includegraphics[height=3in]{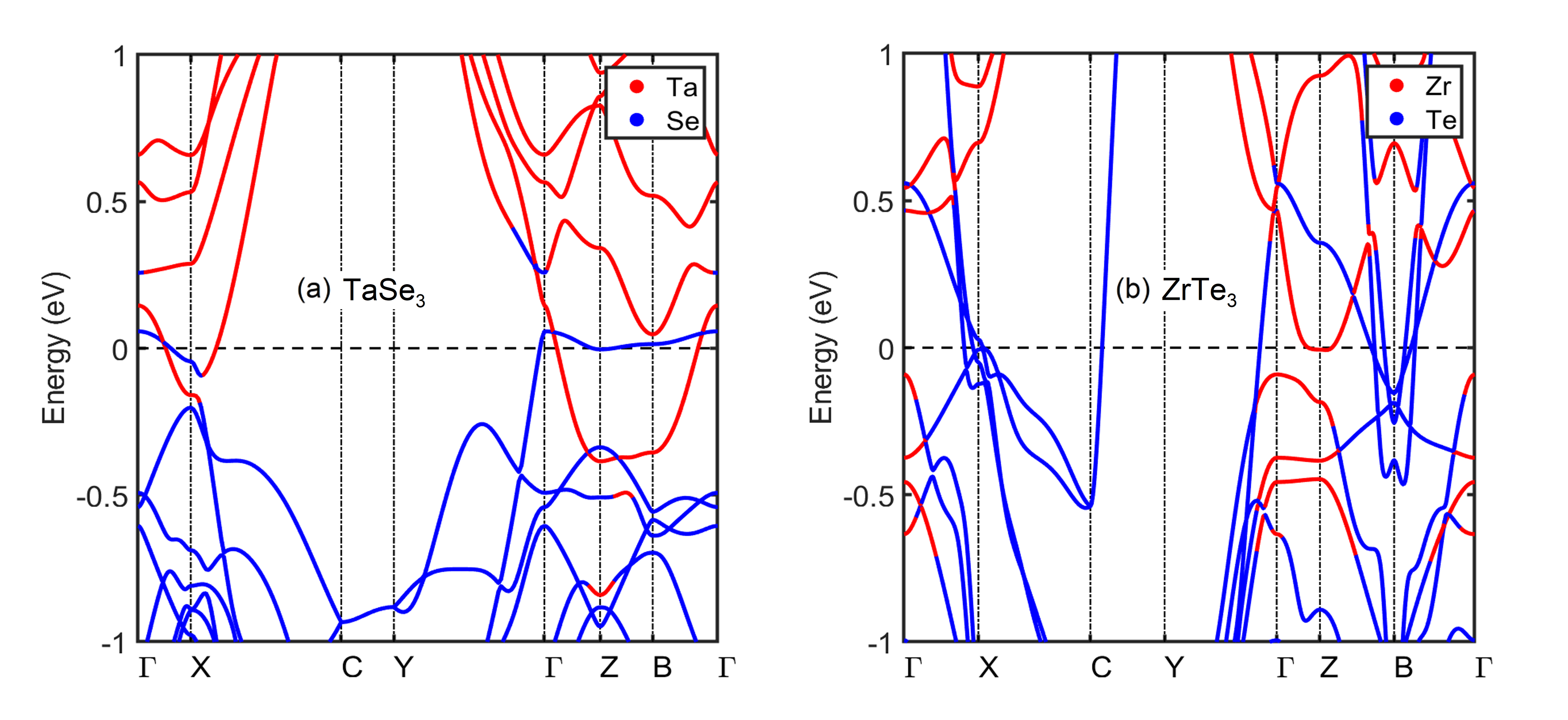}
\caption{Electronic structure of (a) \ta and (b) \zr. The color indicates the dominant orbital contributions from the metal or chalcogen atoms.
}
\label{fig:Ek}
\end{figure*}

\begin{figure*}[!ht]
\centering \includegraphics[height = 2.75in]{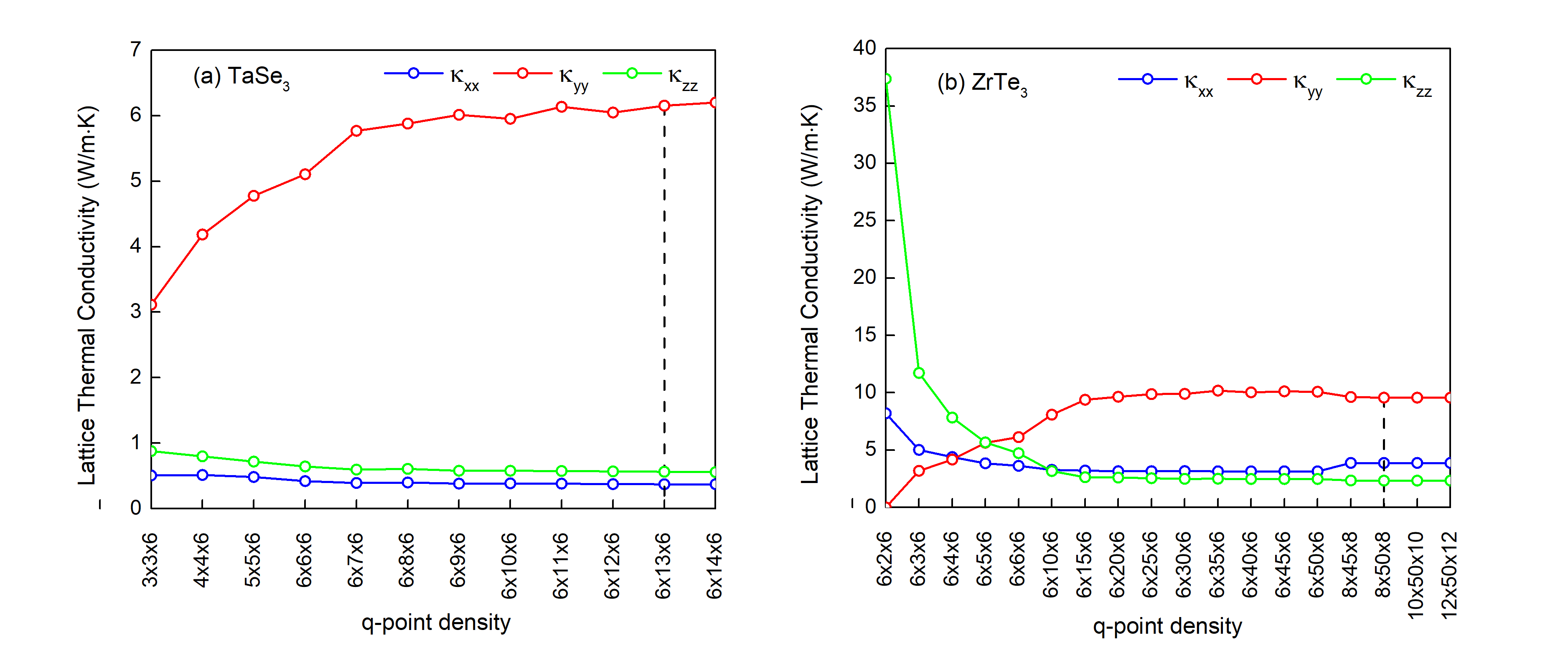}
\caption{
Lattice thermal conductivity for (a) \ta and (b) \zr for different $q$-point grids at T = 300K. 
For our calculation we have chosen a $q$-point of 6$\times$13$\times$6 and 8$\times$50$\times$8 for \ta and \zr, respectively. 
The change of thermal conductivity beyond this q-point is negligible and below 1\%.
}
\label{fig:TC-conv}
\end{figure*}

\clearpage

%

\end{document}